\documentclass[rnote,structabstract]{aa}  
\usepackage{natbib}
\bibpunct{(}{)}{,}{a}{}{,} % to follow the A&A style
\usepackage{graphicx}
\usepackage{txfonts}
\usepackage{dcolumn}
\usepackage{multirow}
\begin{document}
\newcommand{\kmps}{km~s$^{-1}$}
\newcommand{\kmpsb}{km~s$^{-1}$ }
\newcommand{\cdho}{C$^{18}$O}
\newcommand{\cdhob}{C$^{18}$O }
\newcommand{\cdso}{C$^{17}$O}
\newcommand{\cdsob}{C$^{17}$O }
\newcommand{\dzco}{$^{12}$CO}
\newcommand{\dzcob}{$^{12}$CO }
\newcommand{\tzco}{$^{13}$CO}
\newcommand{\tzcob}{$^{13}$CO }
\newcommand{\ndhp}{N$_2$H$^+$}
\newcommand{\ndhpb}{N$_2$H$^+$ }
\newcommand{\nddp}{N$_2$D$^+$}
\newcommand{\nddpb}{N$_2$D$^+$ }
\newcommand{\hdb}{H$_2$ }
\newcommand{\ddb}{D$_2$ }
\newcommand{\htp}{H$_3^+$}
\newcommand{\htpb}{H$_3^+$ }
\newcommand{\dtp}{D$_3^+$}
\newcommand{\dtpb}{D$_3^+$ }
\newcommand{\hddp}{H$_2$D$^+$}
\newcommand{\hddpb}{H$_2$D$^+$ }
\newcommand{\ddhp}{D$_2$H$^+$}
\newcommand{\ddhpb}{D$_2$H$^+$ }
\newcommand{\cd}{column density}
\newcommand{\cdb}{column density }
\newcommand{\cc}{cm$^{-3}$}
\newcommand{\ccb}{cm$^{-3}$ }
\newcommand{\sqc}{cm$^{-2}$}
\newcommand{\sqcb}{cm$^{-2}$ }
\newcommand{\ctds}{C$^{32}$S}
\newcommand{\ctdsb}{C$^{32}$S }
\newcommand{\ctqs}{C$^{34}$S}
\newcommand{\ctqsb}{C$^{34}$S }
\newcommand{\tdso}{$^{32}$SO}
\newcommand{\tdsob}{$^{32}$SO }
\newcommand{\tqso}{$^{34}$SO}
\newcommand{\tqsob}{$^{34}$SO }
\newcommand{\juz}{(J:1--0)}
\newcommand{\juzb}{(J:1--0) }
\newcommand{\jdu}{(J:2--1)}
\newcommand{\jdub}{(J:2--1) }
\newcommand{\jtd}{(J:3--2)}
\newcommand{\jtdb}{(J:3--2) }
\newcommand{\jqt}{(J:4--3)}
\newcommand{\jqtb}{(J:4--3) }
\newcommand{\jcq}{(J:5--4)}
\newcommand{\jcqb}{(J:5--4) }
\newcommand{\jsc}{(J:6--5)}
\newcommand{\jscb}{(J:6--5) }
\newcommand{\jkk}{J$_\mathrm{KK\arcmin}$}
\newcommand{\jkkb}{J$_\mathrm{KK\arcmin}$ }
\newcommand{\jff}{J$_\mathrm{FF\arcmin}$}
\newcommand{\jffb}{J$_\mathrm{FF\arcmin}$ }
\newcommand{\nhdd}{NH$_2$D}
\newcommand{\nhddb}{NH$_2$D }
\newcommand{\den}{n(H$_2$)}
\newcommand{\denb}{n(H$_2$) }
\newcommand{\mjy}{MJy/sr}%$^{-1}$}
\newcommand{\mjyb}{MJy/sr}%$^{-1}$ }
\newcommand{\Av}{A$_{\mathrm V}$}
\newcommand{\Avb}{A$_{\mathrm V}$ }
\newcommand{\SM}{M$_\odot$}
\newcommand{\SMb}{M$_\odot$ }
\newcommand{\pdix}[1]{$\times$ 10$^{#1}$}
\newcommand{\pdixb}[1]{$\times$ 10$^{#1}$ }

   \title{On the frequency of N$_2$H$^+$ and N$_2$D$^+$\thanks{Based on
   observations made with the IRAM 30-m and the GBT 100-m. IRAM is
   supported by INSU/CNRS (France), MPG (Germany), and IGN (Spain). GBT is run by the National Radio Astronomy Observatory which is a facility of the National Science Foundation operated under cooperative agreement by Associated Universities, Inc.}}

%   \subtitle{I. Overviewing the $\kappa$-mechanism}

   \author{L. Pagani
          \inst{1}
          \and
          F. Daniel
          \inst{1,2}
          \and
          M.L. Dubernet
          \inst{1}
          }

    \offprints{L.Pagani}

 \institute{ LERMA \& UMR8112 du CNRS, Observatoire de
  Paris, 61, Av. de l'Observatoire, 75014 Paris, France\\
\email{laurent.pagani@obspm.fr, marie-lise.dubernet@obspm.fr}
\and
Department of Molecular and Infrared Astrophysics ( DAMIR), Consejo 
Superior de Investigaciones Cient{\'i}Þcas (CSIC ), C/ Serrano 121, 28006 Madrid, 
Spain\\
\email{ daniel@damir.iem.csic.es}
}

   \date{received : 11/7/2008; accepted : 10/11/2008}

% \abstract{}{}{}{}{} 
% 5 {} token are mandatory
 
  \abstract
   {Dynamical studies of prestellar cores search for small velocity differences between different tracers. The highest radiation frequency precision is therefore required for each of these species.}
   {We want to adjust the frequency of the first three rotational transitions of \ndhpb and \nddpb and extrapolate to the next three transitions.}
   {\ndhpb and \nddpb are compared to NH$_3$ the frequency of which is more accurately known and which has the advantage to be spatially coexistent with \ndhpb and \nddpb in dark cloud cores. With lines among the narrowests, and \ndhpb and NH$_3$ emitting region among the largests, \object{L183} is a good candidate to compare these species.}
   {A correction of $\sim$10 kHz for the \ndhpb \juzb transition has been found ($\sim$ 0.03Ê\kmps) and similar corrections, from a few m\,s$^{-1}$ up to $\sim$0.05 \kmps are reported for the other transitions (\ndhpb \jtdb and \nddp \juz, \jdu, and \jtd) compared to previous astronomical determinations. Einstein spontaneous decay coefficients (A$_{ul}$) are included.}
   {}


   \maketitle
%
%________________________________________________________________

\section{Introduction}

In the quest for star forming cores, kinematic studies play a crucial role, trying to unveil slowly contracting cores or fast collapsing ones, depending upon which theory we rely upon or at what moment along the evolutionary track the prestellar core is standing. As already discussed by \citet{Lee99}, the accurate knowledge of every species line frequency is of the uttermost importance to track small systematic velocity gradients in molecular clouds. Therefore, because these velocity shifts can be as small as a few tens of m\,s$^{-1}$, millimeter line transitions should be known with a precision of at least 10$^{-7}$ and ideally 10$^{-8}$.
Some species are easily measured in the laboratory, especially stable species like CO, NH$_3$, etc,. Others are unstable and more difficult to measure (such as OH, \hddp,...). One possibility in the latter case, is to compare the transitions of the species of interest with the transitions of another well-known species in dark cloud cores where the lines are narrow enough to be accurately measured. However, the obvious difficulty is to be sure that the two species share the same volume of the cloud and undergo the same macroscopic velocity shifts. Even though, the line opacities might be a problem if too different in presence of a velocity gradient\,: the two coexistent species might then emphasize different parts of the cloud, depending on the depth for which their respective opacity reaches 1. A problem of opacity was indeed met in the comparison of CS with CCS made by \citet{Kuiper96} in their attempt to measure the frequency of the CS lines as discussed in \citet{Pagani01}.

\citet{Caselli95} performed such a measurement for \ndhp, comparing \ndhpb \juzb line emission to the C$_3$H$_2$ (\jkk\,: 2$_{12}$--1$_{01}$) line emission in \object{L1512}, confirming a sizeable difference between laboratory measurement and astronomical observations. \citet{Dore04} expanding on a previous work by \citet{Gerin01} also calculated and observed the \nddpb \juzb transition in \object{L183}, and extrapolated to the higher \nddpb transitions, \citep[giving slightly different values compared to][for the J:2--1 and J:3--2 transitions]{Gerin01}. They aligned their \nddpb \juzb observation onto their \ndhpb \juzb towards the same source with the same telescope. The \ndhpb rotational constant was itself redetermined from a new evaluation of the \ndhpb \juzb frequency from a comparison with \cdhob \juzb in the L1512 cloud \citep[see][for more details]{Dore04}. This new value gave an offset of -4.2 kHz from their previous determination. 

While the direct comparison of the \nddpb and \ndhpb lines is presently the best option to choose because \ndhpb forcibly exists where \nddpb exists, the hypothesis that C$_3$H$_2$ is also present in the same volume as \ndhpb is more questionable because of differential depletion problems. \citet{Dore04} also note that using \cdhob has the problem of tracing different regions but hoped for a null velocity shift between the two tracers. We think that a better possibility exists to measure accurately the frequency of \ndhp, namely by taking NH$_3$ as the frequency reference. NH$_3$ and \ndhpb are clearly coexistent species in depleted prestellar cores \citep[e.g.][]{Tafalla02,Tafalla04}, having a common chemical origin and showing similar extents in most cores. 

In this Note, we present a detailed comparison of NH$_3$ with \ndhpb and \nddpb in L183, checking that the measurable velocity shifts across the core are the same for all three species to convince ourselves of their coexistence and the absence of opacity effect on the velocity peak position.  \citet{Schmidburgk04} developed a similar strategy in their study of H$^{13}$CO$^+$ and \tzcob hyperfine structure (hereafter HFS) towards another dark cloud, \object{L1512}, with similar very narrow linewidths. With these comparisons in hand, we give all corrections for the 5 most currently observed transitions together with their Einstein spontaneous decay coefficients (A$_{ul}$), determine the best fitting rotational constants and compute the expected frequencies for the next 3 rotational transitions (J:4--3, 5--4, 6--5). 
  
%__________________________________________________________________

\section{Observations}

\begin{figure}[tbp]
\centering
\includegraphics[width=11.5cm,angle=-90]{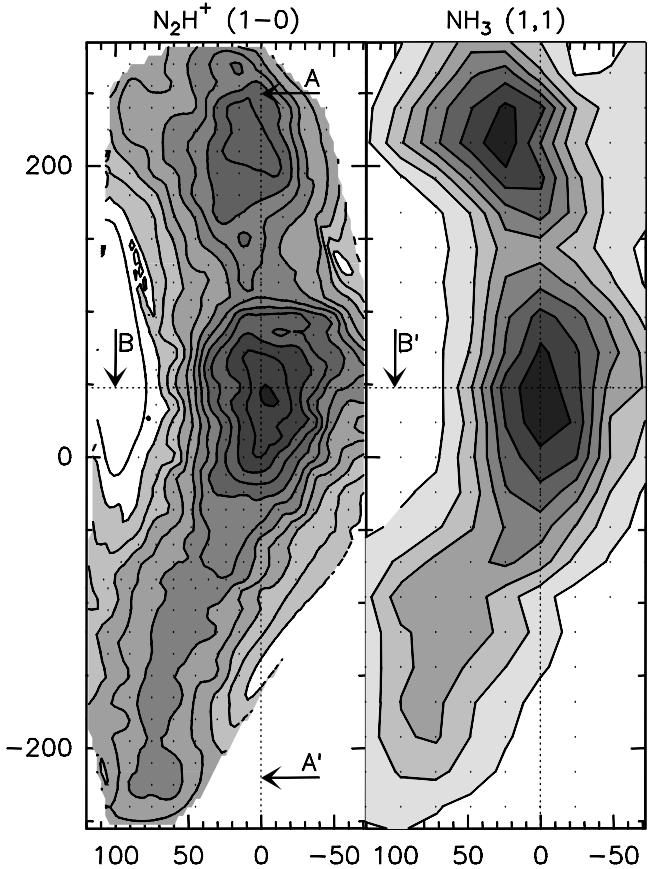}
\caption{\ndhpb \juzb (left) and NH$_3$ (1,1) (right) integrated intensity maps. The dotted lines AA$^\prime$ and BB$^\prime$ indicate the profiles along which the velocity gradients are traced in Figs. \ref{figvelgrada} \& \ref{figvelgradb}. Reference position : $\alpha_{2000}$ = 15$^h$54$^m$08.5$^s$ $\delta_{2000}$ = -2\degr52\arcmin48\arcsec }
\label{fign2hpnh3}
\end{figure}

The whole elongated dense core of L183 (reference position : $\alpha_{2000}$ = 15$^h$54$^m$08.5$^s$ $\delta_{2000}$ = -2\degr52\arcmin48\arcsec) has now been fully mapped wih the IRAM 30-m telescope in a series of observations spanning several years from November 2003 to July 2007. The \ndhpb and \nddpb \juzb lines have been fully mapped while the \ndhpb \jtd, \nddpb \jdub and \jtdb lines have been mapped mostly towards the main core and its elongated ridge and partly towards the peak of the northern core \citep[see][]{Pagani04,Pagani05}. All observations have been performed in frequency--switch mode.  For the \juzb lines, the frequency sampling is 10 kHz, 10 or 20 kHz for the \jdub and 40 kHz for the \jtdb lines, providing comparable velocity resolution for all lines in the range 30--50 m\,s$^{-1}$. Spatial resolution ranges from 33\arcsec at 77 GHz to 9\arcsec at 279 GHz. For all lines, the spatial sampling is 12\arcsec\ for the main prestellar core and 15\arcsec for the southern extension and for the northern prestellar core. We use \citet{Caselli95} and \citet{Dore04} frequencies for \ndhpb and \nddpb transitions, respectively.

We performed observations of NH$_3$ (1,1) and (2,2) inversion lines
towards the whole core at the new Green Bank 100-m
telescope (GBT) in November 2006 and March 2007 with velocity sampling of 20
m\,s$^{-1}$ and a typical T$_\mathrm{sys}$ of 50 K, in frequency--switch mode. The angular resolution ($\sim$35\arcsec) is close to that of the 30-m for the low-frequency (J:1--0) N$_2$D$^+ $line. The spatial sampling is 24\arcsec\ all over the source. We use the accurate measurement of \citet{Kukolich67} for NH$_3$ (1,1), namely $\nu$~=~23\,694\,495\,487 ($\pm$48) Hz which is an average estimated from the whole HFS \citep[see also][who revisited the NH$_3$ and $^{15}$NH$_3$ frequencies. The reported accuracy is higher but the NH$_3$ (1,1) frequency remains basically unchanged, namely $\nu$~=~23\,694\,495\,481 $\pm$ 22 Hz]{Hougen72}. For this frequency, the two strongest hyperfine components have the following frequency offsets\,:\\
\\
$\Delta\nu$(F$_1$F\,: 2,$^5/_2$ $\rightarrow$ 2,$^5/_2$) = 10\,463 Hz \\
\\
$\Delta\nu$(F$_1$F\,: 2,$^3/_2$ $\rightarrow$ 2,$^3/_2$) = -15\,196 Hz \\

Samples of these spectra (\ndhp, \nddpb and NH$_3$) are displayed in \citet{Pagani07}.

\section{Spatial coexistence of ammonia and diazenylium}
\begin{figure*}[tbp]
\centering
\includegraphics[width=11.5cm,angle=-90]{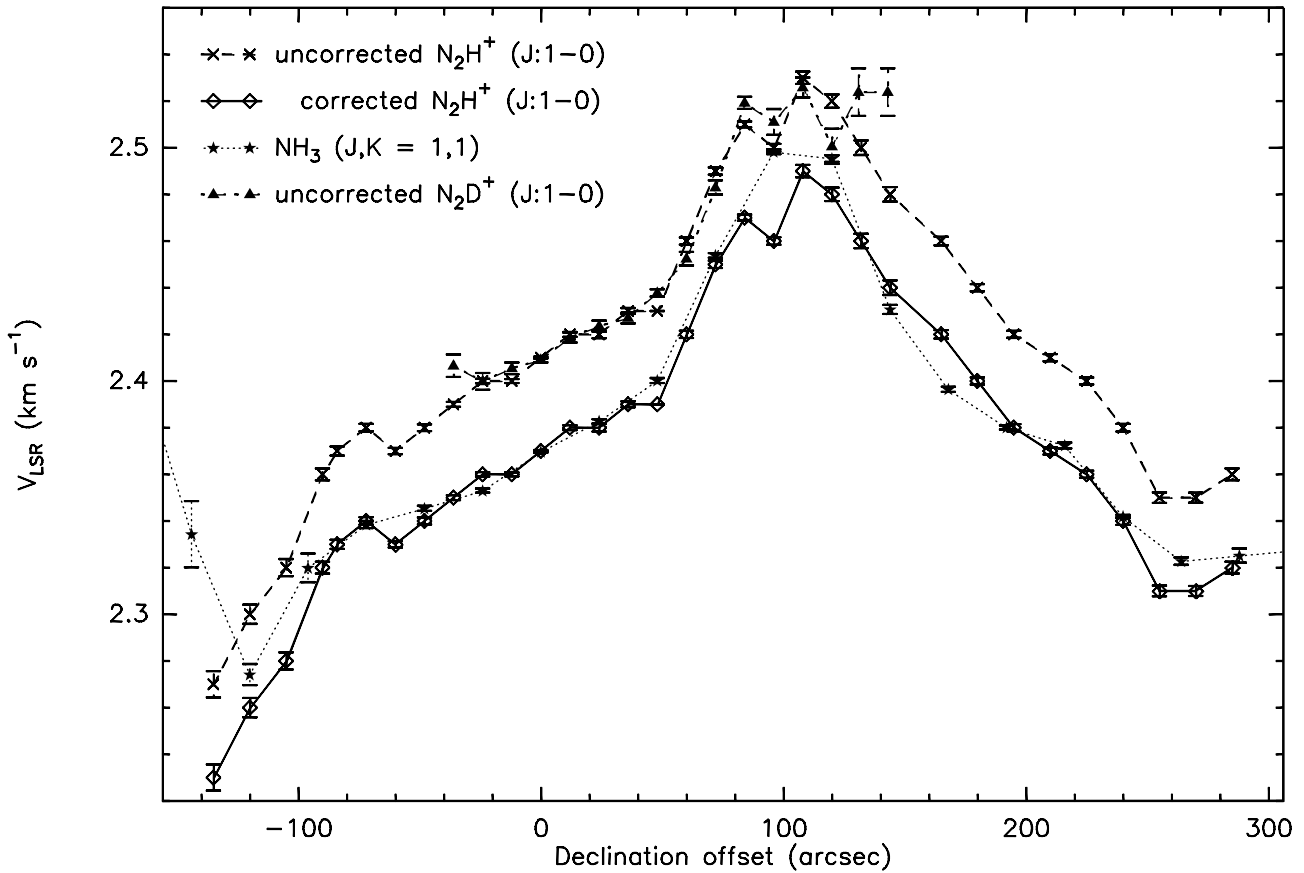}
\caption{\ndhp, \nddpb \juzb and NH$_3$ (1,1) line of sight velocity along the AA$^\prime$ cut (see Fig. \ref{fign2hpnh3}). The \ndhpb data are displayed with the original frequency (uncorrected) and with a correction of -41 m\,s$^{-1}$. The uncorrected \nddpb \juzb points are consistent with the uncorrected \ndhpb points despite the different opacities}
\label{figvelgrada}
\end{figure*}

Though depletion of molecules was predicted in the 70's, it was only a few years after the publication of the \citet{Caselli95} paper on the frequency of \ndhpb that depletion was actually discovered and traced \citep[e.g.][]{Willacy98}. Therefore the hypothesis made by \citet{Caselli95} that C$_3$H$_2$ and \ndhpb are spatially coexistent is probably refutable as it is clear now that such heavy Carbon carrier should be depleted in the same region as CO which is the region where \ndhpb appears. Indeed, the detection of \nddpb in L1512 as a large fraction of \ndhpb \citep{Roberts07} is a clear sign of heavy depletion of other molecules. Therefore the velocity coincidence between these two species is questionable.

\begin{figure}[tbp]
\centering
\includegraphics[width=5.5cm,angle=-90]{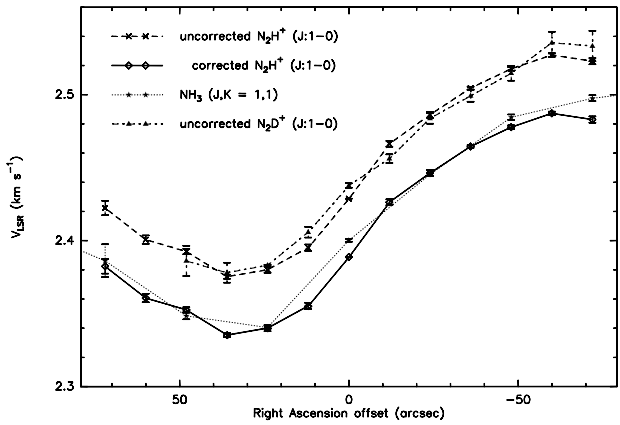}
\caption{ \ndhp, \nddpb \juzb and NH$_3$ (1,1) line of sight velocity along the BB$^\prime$ cut (see Fig. \ref{fign2hpnh3}). The \ndhpb data are displayed with the original frequency (uncorrected) and with a correction of -41 m\,s$^{-1}$. The uncorrected \nddpb \juzb points are consistent with the uncorrected \ndhpb points despite the different opacities}
\label{figvelgradb}
\end{figure}

Ammonia and diazenylium have the same chemical origin, starting from N$_2$ and are well-known to be coexistent as discussed by e.g. \citet{Tafalla02, Tafalla04}. This is in particular true in L183 as can be seen in Fig. \ref{fign2hpnh3} %figure des contours de N2H+ et de NH3 avec les deux coupes en vitesse discutŽes dans le texte
 \citep[but not for C$_3$H$_2$ which has a much smaller extent, mostly concentrated towards the northern prestellar core as can be seen in][]{Swade89}. Interestingly, the velocity along the dense filament is constantly changing (Fig. \ref{figvelgrada}), evoking a flow towards the prestellar cores and the cut perpendicular to the filament (marked BB\arcmin\ in Fig. \ref{fign2hpnh3}) is suggesting a rotation of the filament around its vertical axis (Fig. \ref{figvelgradb}). NH$_3$~(1,1), \ndhpb and \nddpb \juzb all trace exactly the same gradients and it seems therefore compulsory that their velocities be identical as there is no obvious possibility that the velocity gradients be exactly parallel but offset from each other, especially in the probable case of the cylinder rotation. With present \ndhpb \juzb frequency as given by \citet{Caselli95}, there is indeed a clear offset with respect to the NH$_3$ velocity gradient, close to 40 m\,s$^{-1}$ \citep[and to 26 m\,s$^{-1}$ compared to the new value in][]{Dore04}. Note also that \citet{Amano05} have reinterpreted \citet{Caselli95} observations along with new laboratory measurements but are therefore plagued by the velocity difference between \ndhpb and C$_3$H$_2$ which appears to exist in view of the present discrepancy between NH$_3$ and \ndhp. Consequently, their best fit (\#2 of their Table 2) is to be considered cautiously. Finally, the fact that \nddpb velocity centroids are almost identical with those of \ndhpb indicates that the different opacities of the lines are not introducing any measurable bias here (though a very tiny shift is possibly visible in Fig. \ref{figvelgradb} where the \nddpb displacement is symmetrically slightly less than the \ndhpb displacement). 
  
In conclusion, the three species are spatially coexistent and trace the same velocities and one must adjust the frequencies of \ndhpb and \nddpb to that of NH$_3$.
 
\section{Frequency corrections}

\subsection{\ndhpb \juzb correction}

Frequency was measured using the MINIMIZE function in CLASS\footnote{http://www.iram.fr/IRAMFR/GILDAS} with the HFS method for all species (for NH$_3$, the HFS method is similar to the internally built NH3(1,1) method). Because it is easier to deal with velocity offsets in CLASS, especially as we have to compare two species at different frequencies, the measurements have all been made in the velocity scale. Velocity differences are subsequently converted into frequency offsets using the approximate doppler shift formula ($\nu = \nu_0(1 -\frac{\delta v}{c}$), $\delta v$ being the velocity offset, c the celerity of light, $\nu$ and $\nu_0$ the corrected and original frequencies). The HFS method, in order to fit all the hyperfine components individually requires that we provide their list with their relative velocities and relative weights, these parameters not being adjusted during the fit. Therefore, we have used the detailed HFS provided by \citet{Caselli95}, \citet{Dore04} and \citet{Kukolich67}. Since an accurate determination of the hyperfine spectroscopic constants depends only slightly on the adopted rotational constants B and D\footnote{indeed, it can be noted that the HFS splitting is in first approximation identical for both \ndhpb and \nddp despite a large, $\sim$20\% variation in B rotational constant}, we can safely use the previously determined ones.
Doing so, our own determination for the relative velocity offsets between the hyperfine components in the J:1--0 line  agree with \citet{Caselli95} with a typical dispersion of 0.7 kHz. Though this is twice as much as the r.m.s. error on our frequency determination of each individual component ($\sigma \sim$ 0.3 kHz), we find that using their offsets or ours, introduces a negligible difference of 0.13 kHz in the J:1--0 transition frequency determination, which is comparable to the r.m.s. error of the fit (0.12 kHz). We also did not find an improvement on the r.m.s. error of the fit itself. For the \ndhpb and \nddpb transitions, the strongest hyperfine transition was given null velocity offset as it was also the strongest hyperfine transition frequency which was used to tune the receivers. The advantage of a complex and strong HFS is that it lowers the uncertainty on the velocity fit, compared to a single line estimate (fitting individually the \ndhpb J:1--0 lines with independent gaussians, gives errors between 0.85 and 1.2 m\,s$^{-1}$ instead of 0.38 m\,s$^{-1}$ with the global HFS fit for the reference spectrum). 

\begin{figure}[tbp]
\centering
\includegraphics[width=5.5cm,angle=-90]{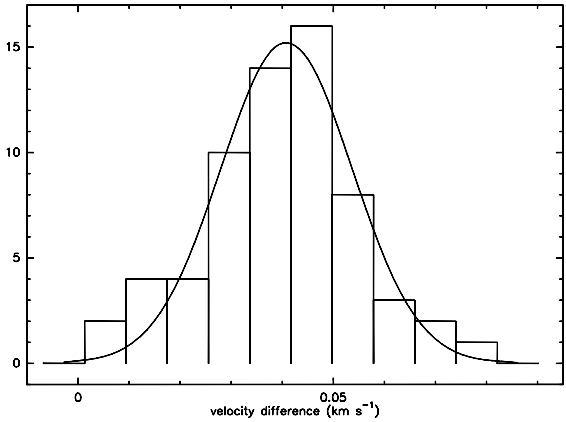}
\caption{ \ndhpb \juzb and NH$_3$ (1,1) line of sight velocity difference histogram. The gaussian fit is centered on 40.8 m\,s$^{-1}$ with a dispersion $\sigma$~=~12.9 m\,s$^{-1}$}
\label{fighisto}
\end{figure}

Though the reference position has been observed often enough to get very high signal-to-noise ratios for most transitions, it seems more secure to measure the offset between \ndhpb and NH$_3$ on all common positions (every other position in the central core, a few positions in the rest of the cloud) and measure the average difference. We have identified 65 common positions with sufficient signal-to-noise ratios and we have obtained the dispersion histogram of the velocity difference (Fig. \ref{fighisto}). Fitting the histogram with a gaussian, we find a velocity difference of 40.8 m\,s$^{-1}$ with a dispersion  $\sigma$~=~12.9~m~s$^{-1}$. This corresponds to a frequency correction of -13 $\pm$ 4 kHz \citep[or -8.8 kHz compared to][]{Dore04}. For the reference position alone, the difference is also 40.8 m\,s$^{-1}$ with an error  $\sigma$~=~0.56~m~s$^{-1}$ (due to the very high signal to noise ratio obtained for both lines towards that position). 

\subsection{\ndhpb \jtdb correction}
For the \ndhpb \jtdb transition, only the reference position has been observed with a reasonably good signal-to-noise ratio ($\sim$10). Therefore, we can only make a direct comparison for this position. The Jet Propulsion Laboratory (JPL) catalogue frequency for this line (279\,511.701 $\pm$ 0.05 MHz) is too vague to be useful for a precise velocity determination. The Cologne Database for Molecular Spectroscopy (CDMS) catalogue gives $\nu$ = 279\,511.8577 MHz for the (F$_1$F: 4,5--3,4) strongest hyperfine component based on various works while \citet{Crapsi05} give 279\,511.863 MHz determined from the new rotational and centrifugal distorsion constants from \citet{Dore04}. These new values are respectively 26 and 31 kHz above our own determination.

\subsection{\nddpb corrections}
For all three transitions of \nddp, we took advantage of the similar sampling with \ndhpb \juzb to have a larger number of comparison points.  We obtained 83, 73 and 51 comparison points with sufficient signal-to-noise ratio between \ndhpb \juzb \citep[using][frequency]{Caselli95} and \nddpb \juz, \jdu, and \jtdb transitions respectively. The gaussian fit to each histogram yielded\,:\\
\\
\juz\,: -5.1 m\,s$^{-1}$ ($\sigma$ = 10.5 m\,s$^{-1}$)\\
\jdu\,: 12.5 m\,s$^{-1}$ ($\sigma$ = 14.4 m\,s$^{-1}$)\\
\jtd\,: 18.4 m\,s$^{-1}$ ($\sigma$ = 8.1 m\,s$^{-1}$)\\
\\
The corresponding correction with respect to NH$_3$(1,1) is\,:\\
\\
\juz\,: 35.7 m\,s$^{-1}$  or -9.2 ($\pm$ 2.7) kHz\\
\jdu\,: 53.3 m\,s$^{-1}$ or -27 ($\pm$ 7.4) kHz\\
\jtd\,: 59.2 m\,s$^{-1}$ or -49 ($\pm$ 6.7) kHz \\
\\
Direct comparison of the reference position with NH$_3$(1,1) spectrum yields\,:\\
\\
\juz\,: 37.7 m\,s$^{-1}$  ($\sigma$ = 0.85 m\,s$^{-1}$)\\
\jdu\,: 47.7 m\,s$^{-1}$ ($\sigma$ = 0.92 m\,s$^{-1}$)\\
\jtd\,: 63.6 m\,s$^{-1}$ ($\sigma$ = 4.7 m\,s$^{-1}$) \\

\subsection{Rotational constants and Einstein--A coefficients}

Except for the \ndhpb \jtdb line which has only one measurement, we have used the averaged comparisons for correcting the frequencies of all these transitions.

We have derived from these new frequencies the rotation (B) and centrifugal distortion (D) constants for \ndhpb and \nddp, using the hyperfine constants given by \citet{Caselli95} and \citet{Dore04} respectively. The error budget has been estimated by adding 1 $\sigma$ to one of the frequency measurements and subtracting 1 $\sigma$ to the other which we use to determine B and D, e.g. +2.7 kHz to the \nddpb \juzb line and -6.7 kHz for the \nddpb \jtdb line. For the \ndhpb \jtdb transition, as we have only one measurement, we have taken the average of the 1$\sigma$ dispersion for all the other transition measurements as a probable dispersion for that measurement if we had had as many observations. We have found an average velocity dispersion of 11.5\,m\,s$^{-1}$ which corresponds to 10.7 kHz at that frequency. The new constants are listed in Table \ref{tab_rotcst}. As expected from the fact that \citet{Amano05} make use of the \citet{Caselli95} frequency determination of \ndhpb and the related \citet{Dore04} \nddpb measurements, their rotational constants are different from ours by an amount directly related to the difference between C$_3$H$_2$ and NH$_3$ velocity determinations. The difference (5.3 kHz for B(\ndhp) and 9.2 kHz for B(\nddp)) is significantly larger than the error estimate (conservatively given to be 2.5 and 1.7 kHz respectively for us and 1.3 and 1.2 kHz for \citealt{Amano05}). It would be interesting to repeat \citet{Amano05} analysis with our new frequency determinations to better secure these values.

Line strengths, from which Einstein--A coefficients are defined, are determined from the 
reduced transition matrix elements of the dipole moment operator:
\begin{eqnarray}
S(1 \to 2) = | \left< \psi_1 || \hat{d} || \psi_2 \right> |^2 
\end{eqnarray}
where $|\psi_1>$ and $|\psi_2>$ are the wave--functions of the two levels involved
in the radiative transition. In the case of hyperfine structures, the wave--functions
can be defined according to an expansion on Hund's case (b) wave--functions, the coefficients being
determined by diagonalisation of the hyperfine Hamiltonian.
In the case of N$_2$H$^+$ and N$_2$D$^+$, the mixing of states is low so that a given 
hyperfine wave--function can be accuretaley defined as a pure Hund's case (b)
wave--function. Doing so, the line strengths can be expressed in a closed form 
\citep{gordy1984}, and for N$_2$H$^+$, the relevant expressions being given in \citet{Daniel06}. 
The Einstein--A coefficients are then given by:
\begin{eqnarray}
A_{J F_1 F \to J' F_1' F'} = \frac{64 \pi^4}{3 h c^3} \mu^2 
\nu^3_{J F_1 F \to J' F_1' F'} \times \frac{J}{[F]} \, s_{J F_1 F \to J' F_1' F'}
\end{eqnarray}
\citep[this is the same equation as in][but corrected for two typos]{Daniel06}

The calculated line frequencies and A$_{ul}$ coefficients (the dipole moment -- $\mu$ = 3.37 D -- is taken from \citealp{Botschwina84}) are given in Tables \ref{tabn2hp10} to \ref{tabn2dp32} for all rotational transitions from \juzb to \jscb for both \ndhpb and \nddp. The frequency uncertainty is estimated by varying the rotational B and D constants by $\pm$1 $\sigma$. 

\begin{table}[htdp]
\caption{Rotation (B) and centrifugal distortion (D) constants for \ndhpb and \nddp. Errors in parentheses are given for the last two digits}
\centering
\begin{tabular}{ccc}
\hline
Species&B&D\\
&MHz&MHz\\
\hline
\ndhp& 46586.8713(25)&0.08796(24) \\
\nddp& 38554.7479(17)&0.06181(15)\\
\hline
\end{tabular}
\label{tab_rotcst}
\end{table}%

\section{Conclusions}
\begin{enumerate}
\item New, more accurate rotational constants and line frequencies are given along with the detailed Einstein spontaneous coefficients (A$_{ul}$) for each of the hyperfine components.
\item The main prestellar core LSR velocity is 2.3670 ($\pm$0.0004) \kmps.
\end{enumerate}

\begin{acknowledgements}
We thank an anonymous referee for her/his critical reading which helped to improve the manuscript.
\end{acknowledgements}

\addtocounter{table}{12}

\onltab{2}{
\begin{table}[htbp]
 \caption{Hyperfine components and A$_{ul}$ Einstein spontaneous emission coefficients of the \juzb transition of N$_2$H$^+$. The frequency uncertainty is $\pm$ 4.0 kHz for all hyperfine components. Summing  A$_{ul}$ over all the hyperfine components with the same frequency always give the same total A$_{ul}$, 3.628 \pdix{-5} s$^{-1}$. 3.628(-5) means 3.628\pdix{-5}}\label{tabn2hp10}.
 \centering
 \begin{tabular}{ccccccD{.}{.}{4}D{.}{.}{3}}
% \begin{tabular}{ccc @{$\rightarrow$} cccD{.}{.}{4}D{.}{.}{3}}
 \hline \noalign{\smallskip}
  $J'$ & $F_1'$ & \multicolumn{2}{c}{$F' \rightarrow J$} & $F_1$ & $F$ & \multicolumn{1}{c}{Frequency} & \multicolumn{1}{c}{A$_{ul}$} \cr
  && \multicolumn{2}{c}{}&&&  \multicolumn{1}{c}{(MHz)}& \multicolumn{1}{c}{(s$^{-1}$)}\cr
 \noalign{\smallskip} \hline \noalign{\smallskip}
1&1&0&0&1&1&93\,171.6081&3.628(-5)\\
1&1&2&0&1&2&93\,171.9049&2.721(-5) \\
1&1&2&0&1&1&93\,171.9049&9.069(-6)\\%&\raisebox{1.5ex}[0pt]{\hspace{0.2cm}\rule[-8pt]{1pt}{18pt}\hspace{0.1cm}3.628(-5)}\\
1&1&1&0&1&0&93\,172.0398&1.209(-5)\\
1&1&1&0&1&2&93\,172.0398&1.512(-5)\\
1&1&1&0&1&1&93\,172.0398&9.069(-6)\\%&\raisebox{2.25ex}[0pt]{\hspace{0.2cm}\rule[-11pt]{1pt}{27pt}\hspace{0.1cm}3.628(-5)}\\
1&2&2&0&1&1&93\,173.4669&2.721(-5)\\
1&2&2&0&1&2&93\,173.4669&9.070(-6)\\%&\raisebox{1.5ex}[0pt]{\hspace{0.2cm}\rule[-8pt]{1pt}{18pt}\hspace{0.1cm}3.628(-5)}\\
1&2&3&0&1&2&93\,173.7637&3.628(-5)\\
1&2&1&0&1&2&93\,173.9540&1.008(-6)\\
1&2&1&0&1&1&93\,173.9540&1.512(-5)\\
1&2&1&0&1&0&93\,173.9540&2.016(-5)\\%&\raisebox{2.25ex}[0pt]{\hspace{0.2cm}\rule[-11pt]{1pt}{27pt}\hspace{0.1cm}3.628(-5)}\\
1&0&1&0&1&1&93\,176.2522&1.209(-5)\\
1&0&1&0&1&2&93\,176.2522&2.016(-5)\\
1&0&1&0&1&0&93\,176.2522&4.031(-6)\\%&\raisebox{2.25ex}[0pt]{\hspace{0.2cm}\rule[-11pt]{1pt}{27pt}\hspace{0.1cm}3.628(-5)}\\
 \noalign{\smallskip} \hline \noalign{\smallskip}
\end{tabular}
\end{table}
}
\onltab{3}{
\begin{table}[htbp]
 \caption{Hyperfine components and A$_{ul}$ Einstein spontaneous emission coefficients of the \jdub transition of N$_2$H$^+$. The frequency uncertainty is $\pm$ 2.3 kHz for all hyperfine components. }\label{tabn2hp21}
 \centering
 \begin{tabular}{ccccccD{.}{.}{4}D{.}{.}{3}}

 \hline \noalign{\smallskip}
  $J'$ & $F_1'$ &  \multicolumn{2}{c}{$F' \rightarrow J$}  & $F_1$ & $F$ & \multicolumn{1}{c}{Frequency} & \multicolumn{1}{c}{A$_{ul}$} \cr
  &&&&&& \multicolumn{1}{c}{(MHz)}& \multicolumn{1}{c}{(s$^{-1}$)}\cr
 \noalign{\smallskip} \hline \noalign{\smallskip}
2&2&2&1&2&1&186\,342.4666& 1.306(-5)\\
2&2&2&1&2&3&186\,342.6570& 1.354(-5)\\
2&2&1&1&2&1&186\,342.7883& 6.530(-5)\\
2&2&3&1&2&3&186\,342.9123& 7.739(-5)\\
2&2&2&1&2&2&186\,342.9537& 6.046(-5)\\
2&1&1&1&0&1&186\,343.0459& 1.935(-4)\\
2&2&3&1&2&2&186\,343.2091& 9.674(-6)\\
2&1&2&1&0&1&186\,343.2577& 1.935(-4)\\
2&2&1&1&2&2&186\,343.2755& 2.177(-5)\\
2&1&0&1&0&1&186\,343.5098& 1.935(-4)\\
2&2&2&1&1&1&186\,344.3808& 1.959(-4)\\
2&3&3&1&2&3&186\,344.4444& 3.870(-5)\\
2&2&2&1&1&2&186\,344.5158& 6.530(-5)\\
2&2&1&1&1&1&186\,344.7026& 1.088(-4)\\
2&3&3&1&2&2&186\,344.7412& 3.096(-4)\\
2&3&2&1&2&1&186\,344.7615& 2.925(-4)\\
2&2&3&1&1&2&186\,344.7711& 2.612(-4)\\
2&2&1&1&1&2&186\,344.8375& 7.255(-6)\\
2&3&4&1&2&3&186\,344.8419& 3.483(-4)\\
2&3&2&1&2&3&186\,344.9519& 1.548(-6)\\
2&2&1&1&1&0&186\,345.1343& 1.451(-4)\\
2&3&2&1&2&2&186\,345.2487& 5.417(-5)\\
2&1&1&1&2&1&186\,345.3441& 2.418(-6)\\
2&1&2&1&2&1&186\,345.5559& 9.674(-8)\\
2&1&2&1&2&3&186\,345.7462& 8.126(-6)\\
2&1&0&1&2&1&186\,345.8080& 9.674(-6)\\
2&1&1&1&2&2&186\,345.8312& 7.256(-6)\\
2&1&2&1&2&2&186\,346.0430& 1.451(-6)\\
2&1&1&1&1&1&186\,347.2584& 3.628(-5)\\
2&1&1&1&1&2&186\,347.3933& 6.046(-5)\\
2&1&2&1&1&1&186\,347.4701& 3.628(-5)\\
2&1&2&1&1&2&186\,347.6050& 1.088(-4)\\
2&1&1&1&1&0&186\,347.6901& 4.837(-5)\\
2&1&0&1&1&1&186\,347.7222& 1.451(-4)\\

 \noalign{\smallskip} \hline \noalign{\smallskip}
\end{tabular}
\end{table}
}
\onltab{4}{
\begin{table}[htbp]
 \caption{Hyperfine components and A$_{ul}$ Einstein spontaneous emission coefficients of the \jtdb transition of N$_2$H$^+$. The frequency uncertainty is $\pm$ 11 kHz for all hyperfine components. }\label{tabn2hp32}
 \centering
 \begin{tabular}{ccccccD{.}{.}{4}D{.}{.}{3}}

 \hline \noalign{\smallskip}
  $J'$ & $F_1'$ &  \multicolumn{2}{c}{$F' \rightarrow J$}  & $F_1$ & $F$ & \multicolumn{1}{c}{Frequency} & \multicolumn{1}{c}{A$_{ul}$} \cr
  &&&&&& \multicolumn{1}{c}{(MHz)}& \multicolumn{1}{c}{(s$^{-1}$)}\cr
 \noalign{\smallskip} \hline \noalign{\smallskip}
3&3&3&2&3&2&279\,509.361&1.110(-5)\\
3&3&3&2&3&4&279\,509.471&1.124(-5)\\
3&3&2&2&3&2&279\,509.795&1.244(-4)\\
3&3&4&2&3&4&279\,509.849&1.312(-4)\\
3&3&3&2&3&3&279\,509.868&1.176(-4)\\
3&3&4&2&3&3&279\,510.246&8.745(-6)\\
3&3&2&2&3&3&279\,510.302&1.555(-5)\\
3&2&2&2&1&2&279\,511.103&2.644(-4)\\
3&2&2&2&1&1&279\,511.315&7.933(-4)\\
3&2&1&2&1&0&279\,511.355&5.876(-4)\\
3&4&4&2&3&4&279\,511.384&7.870(-5)\\
3&3&3&2&2&3&279\,511.401&1.244(-4)\\
3&2&3&2&1&2&279\,511.479&1.058(-3)\\
3&2&1&2&1&2&279\,511.607&2.938(-5)\\
3&3&3&2&2&2&279\,511.656&9.950(-4)\\
3&3&2&2&2&1&279\,511.768&9.402(-4)\\
3&3&4&2&2&3&279\,511.778&1.119(-3)\\
3&4&3&2&3&2&279\,511.780&1.156(-3)\\
3&4&4&2&3&3&279\,511.781&1.181(-3)\\
3&2&1&2&1&1&279\,511.819&4.407(-4)\\
3&4&5&2&3&4&279\,511.832&1.259(-3)\\
3&3&2&2&2&3&279\,511.834&4.975(-6)\\
3&4&3&2&3&4&279\,511.890&1.606(-6)\\
3&2&2&2&3&2&279\,511.897&6.218(-7)\\
3&3&2&2&2&2&279\,512.090&1.741(-4)\\
3&2&3&2&3&2&279\,512.273&1.269(-8)\\
3&4&3&2&3&3&279\,512.287&1.012(-4)\\
3&2&3&2&3&4&279\,512.383&5.140(-6)\\
3&2&1&2&3&2&279\,512.401&5.597(-6)\\
3&2&2&2&3&3&279\,512.405&4.975(-6)\\
3&2&3&2&3&3&279\,512.781&4.442(-7)\\
3&2&2&2&2&1&279\,513.870&2.938(-5)\\
3&2&2&2&2&3&279\,513.937&3.047(-5)\\
3&2&2&2&2&2&279\,514.192&1.360(-4)\\
3&2&3&2&2&3&279\,514.313&1.741(-4)\\
3&2&1&2&2&1&279\,514.374&1.469(-4)\\
3&2&3&2&2&2&279\,514.568&2.177(-5)\\
3&2&1&2&2&2&279\,514.696&4.897(-5)\\
 \noalign{\smallskip} \hline \noalign{\smallskip}
\end{tabular}
\end{table}
}
\onltab{5}{
\begin{table}[htbp]
 \caption{Hyperfine components and A$_{ul}$ Einstein spontaneous emission coefficients of the \jqtb transition of N$_2$H$^+$. The frequency uncertainty is $\pm$ 41 kHz for all hyperfine components. }\label{tabn2hp43}
 \centering
 \begin{tabular}{ccccccD{.}{.}{4}D{.}{.}{3}}

 \hline \noalign{\smallskip}
  $J'$ & $F_1'$ &  \multicolumn{2}{c}{$F' \rightarrow J$}  & $F_1$ & $F$ & \multicolumn{1}{c}{Frequency} & \multicolumn{1}{c}{A$_{ul}$} \cr
  &&&&&& \multicolumn{1}{c}{(MHz)}& \multicolumn{1}{c}{(s$^{-1}$)}\cr
 \noalign{\smallskip} \hline \noalign{\smallskip}
4&4&4&3&4&3&372\,670.005& 9.404(-6)\\
4&4&4&3&4&5&372\,670.062& 9.457(-6)\\
4&4&3&3&4&3&372\,670.467& 1.814(-4)\\
4&4&5&3&4&5&372\,670.500& 1.857(-4)\\
4&4&4&3&4&4&372\,670.510& 1.746(-4)\\
4&4&5&3&4&4&372\,670.948& 7.738(-6)\\
4&4&3&3&4&4&372\,670.972& 1.209(-5)\\
4&3&3&3&2&3&372\,671.904& 3.158(-4)\\
4&5&5&3&4&5&372\,672.046& 1.238(-4)\\
4&4&4&3&3&4&372\,672.046& 1.814(-4)\\
4&3&3&3&2&2&372\,672.280& 2.527(-3)\\
4&3&2&3&2&1&372\,672.280& 2.388(-3)\\
4&3&4&3&2&3&372\,672.348& 2.843(-3)\\
4&3&3&3&4&3&372\,672.398& 2.467(-7)\\
4&3&2&3&2&3&372\,672.408& 1.263(-5)\\
4&4&4&3&3&3&372\,672.423& 2.720(-3)\\
4&4&3&3&3&2&372\,672.452& 2.665(-3)\\
4&4&5&3&3&4&372\,672.484& 2.902(-3)\\
4&5&4&3&4&3&372\,672.486& 2.942(-3)\\
4&5&5&3&4&4&372\,672.494& 2.971(-3)\\
4&4&3&3&3&4&372\,672.508& 3.701(-6)\\
4&5&6&3&4&5&372\,672.526& 3.095(-3)\\
4&5&4&3&4&5&372\,672.544& 1.528(-6)\\
4&3&2&3&2&2&372\,672.784& 4.422(-4)\\
4&3&4&3&4&3&372\,672.842& 3.046(-9)\\
4&4&3&3&3&3&372\,672.885& 2.332(-4)\\
4&3&4&3&4&5&372\,672.899& 3.753(-6)\\
4&3&2&3&4&3&372\,672.902& 3.948(-6)\\
4&3&3&3&4&4&372\,672.903& 3.701(-6)\\
4&5&4&3&4&4&372\,672.992& 1.513(-4)\\
4&3&4&3&4&4&372\,673.347& 1.919(-7)\\
4&3&3&3&3&2&372\,674.383& 1.974(-5)\\
4&3&3&3&3&4&372\,674.439& 1.999(-5)\\
4&3&3&3&3&3&372\,674.816& 2.090(-4)\\
4&3&4&3&3&4&372\,674.883& 2.332(-4)\\
4&3&2&3&3&2&372\,674.887& 2.211(-4)\\
4&3&4&3&3&3&372\,675.260& 1.555(-5)\\
4&3&2&3&3&3&372\,675.320& 2.764(-5)\\
 \noalign{\smallskip} \hline \noalign{\smallskip}
\end{tabular}
\end{table}
}
\onltab{6}{
\begin{table}[htbp]
 \caption{Hyperfine components and A$_{ul}$ Einstein spontaneous emission coefficients of the \jcqb transition of N$_2$H$^+$. The frequency uncertainty is $\pm$ 95 kHz for all hyperfine components. }\label{tabn2hp54}
 \centering
 \begin{tabular}{ccccccD{.}{.}{4}D{.}{.}{3}}

 \hline \noalign{\smallskip}
  $J'$ & $F_1'$ &  \multicolumn{2}{c}{$F' \rightarrow J$}  & $F_1$ & $F$ & \multicolumn{1}{c}{Frequency} & \multicolumn{1}{c}{A$_{ul}$} \cr
  &&&&&& \multicolumn{1}{c}{(MHz)}& \multicolumn{1}{c}{(s$^{-1}$)}\cr
 \noalign{\smallskip} \hline \noalign{\smallskip}
5&5&5&4&5&4&465\,822.236& 8.093(-6)\\
5&5&5&4&5&6&465\,822.254& 8.118(-6)\\
5&5&4&4&5&4&465\,822.704& 2.374(-4)\\
5&5&6&4&5&6&465\,822.729& 2.404(-4)\\
5&5&5&4&5&5&465\,822.734& 2.311(-4)\\
5&5&4&4&5&5&465\,823.202& 9.891(-6)\\
5&5&6&4&5&5&465\,823.209& 6.869(-6)\\
5&4&4&4&3&4&465\,824.191& 3.673(-4)\\
5&5&5&4&4&5&465\,824.279& 2.374(-4)\\
5&6&6&4&5&6&465\,824.285& 1.717(-4)\\
5&4&4&4&5&4&465\,824.546& 1.221(-7)\\
5&4&3&4&3&2&465\,824.627& 5.397(-3)\\
5&4&4&4&3&3&465\,824.635& 5.509(-3)\\
5&4&5&4&3&4&465\,824.673& 5.877(-3)\\
5&4&3&4&3&4&465\,824.687& 7.496(-6)\\
5&5&5&4&4&4&465\,824.717& 5.697(-3)\\
5&5&4&4&4&3&465\,824.723& 5.642(-3)\\
5&5&4&4&4&5&465\,824.747& 2.931(-6)\\
5&5&6&4&4&5&465\,824.754& 5.935(-3)\\
5&6&5&4&5&4&465\,824.756& 5.978(-3)\\
5&6&6&4&5&5&465\,824.765& 6.010(-3)\\
5&6&5&4&5&6&465\,824.774& 1.419(-6)\\
5&6&7&4&5&6&465\,824.788& 6.182(-3)\\
5&4&5&4&5&4&465\,825.029& 1.009(-9)\\
5&4&3&4&5&4&465\,825.042& 3.053(-6)\\
5&4&4&4&5&5&465\,825.045& 2.931(-6)\\
5&4&5&4&5&6&465\,825.047& 2.952(-6)\\
5&4&3&4&3&3&465\,825.131& 4.722(-4)\\
5&5&4&4&4&4&465\,825.185& 2.901(-4)\\
5&6&5&4&5&5&465\,825.254& 2.029(-4)\\
5&4&5&4&5&5&465\,825.527& 9.991(-8)\\
5&4&4&4&4&3&465\,826.566& 1.469(-5)\\
5&4&4&4&4&5&465\,826.590& 1.478(-5)\\
5&4&4&4&4&4&465\,827.028& 2.728(-4)\\
5&4&3&4&4&3&465\,827.062& 2.833(-4)\\
5&4&5&4&4&5&465\,827.072& 2.902(-4)\\
5&4&5&4&4&4&465\,827.510& 1.209(-5)\\
5&4&3&4&4&4&465\,827.524& 1.889(-5)\\
 \noalign{\smallskip} \hline \noalign{\smallskip}
\end{tabular}
\end{table}
}
\onltab{7}{
\begin{table}[htbp]
 \caption{Hyperfine components and A$_{ul}$ Einstein spontaneous emission coefficients of the \jscb transition of N$_2$H$^+$. The frequency uncertainty is $\pm$ 0.18 MHz for all hyperfine components. }\label{tabn2hp65}
 \centering
 \begin{tabular}{ccccccD{.}{.}{4}D{.}{.}{3}}

 \hline \noalign{\smallskip}
  $J'$ & $F_1'$ &  \multicolumn{2}{c}{$F' \rightarrow J$}  & $F_1$ & $F$ & \multicolumn{1}{c}{Frequency} & \multicolumn{1}{c}{A$_{ul}$} \cr
  &&&&&& \multicolumn{1}{c}{(MHz)}& \multicolumn{1}{c}{(s$^{-1}$)}\cr
 \noalign{\smallskip} \hline \noalign{\smallskip}
6&6&6&5&6&7&558\,963.91& 7.094(-6)\\
6&6&6&5&6&5&558\,963.93& 7.081(-6)\\
6&6&5&5&6&5&558\,964.39& 2.929(-4)\\
6&6&7&5&6&7&558\,964.41& 2.951(-4)\\
6&6&6&5&6&6&558\,964.42& 2.871(-4)\\
6&6&5&5&6&6&558\,964.88& 8.368(-6)\\
6&6&7&5&6&6&558\,964.92& 6.148(-6)\\
6&5&5&5&4&5&558\,965.91& 4.195(-4)\\
6&6&6&5&5&6&558\,965.97& 2.929(-4)\\
6&7&7&5&6&7&558\,965.98& 2.213(-4)\\
6&5&5&5&6&5&558\,966.18& 6.916(-8)\\
6&5&4&5&4&3&558\,966.38& 9.969(-3)\\
6&5&5&5&4&4&558\,966.39& 1.007(-2)\\
6&5&4&5&4&5&558\,966.39& 5.179(-6)\\
6&5&6&5&4&5&558\,966.42& 1.049(-2)\\
6&6&5&5&5&6&558\,966.44& 2.421(-6)\\
6&6&5&5&5&4&558\,966.44& 1.020(-2)\\
6&6&6&5&5&5&558\,966.45& 1.025(-2)\\
6&7&6&5&6&7&558\,966.46& 1.310(-6)\\
6&6&7&5&5&6&558\,966.47& 1.054(-2)\\
6&7&6&5&6&5&558\,966.47& 1.059(-2)\\
6&7&7&5&6&6&558\,966.48& 1.062(-2)\\
6&7&8&5&6&7&558\,966.50& 1.085(-2)\\
6&5&4&5&6&5&558\,966.67& 2.490(-6)\\
6&5&5&5&6&6&558\,966.67& 2.421(-6)\\
6&5&6&5&6&7&558\,966.68& 2.431(-6)\\
6&5&6&5&6&5&558\,966.69& 4.092e-10\\
6&5&4&5&4&4&558\,966.88& 5.127(-4)\\
6&6&5&5&5&5&558\,966.91& 3.462(-4)\\
6&7&6&5&6&6&558\,966.96& 2.554(-4)\\
6&5&6&5&6&6&558\,967.18& 5.852(-8)\\
6&5&5&5&5&6&558\,968.27& 1.169(-5)\\
6&5&5&5&5&4&558\,968.23& 1.165(-5)\\
6&5&5&5&5&5&558\,968.70& 3.327(-4)\\
6&5&4&5&5&4&558\,968.72& 3.418(-4)\\
6&5&6&5&5&6&558\,968.74& 3.462(-4)\\
6&5&4&5&5&5&558\,969.19& 1.424(-5)\\
6&5&6&5&5&5&558\,969.21& 9.890(-6)\\
 \noalign{\smallskip} \hline \noalign{\smallskip}
\end{tabular}
\end{table}
}
\onltab{8}{
\begin{table}[htbp]
\vspace{1cm}
 \caption{Hyperfine components and A$_{ul}$ Einstein spontaneous emission coefficients of the \juzb  transition of N$_2$D$^+$. The frequency uncertainty is $\pm$ 2.8 kHz for all hyperfine components. }\label{tabn2dp10}
 \centering
 \begin{tabular}{ccccccD{.}{.}{4}D{.}{.}{3}}

 \hline \noalign{\smallskip}
  $J'$ & $F_1'$ & \multicolumn{2}{c}{$F' \rightarrow J$} & $F_1$ & $F$ & \multicolumn{1}{c}{Frequency} & \multicolumn{1}{c}{A$_{ul}$} \cr
  &&&&&& \multicolumn{1}{c}{(MHz)}& \multicolumn{1}{c}{(s$^{-1}$)}\cr
 \noalign{\smallskip} \hline \noalign{\smallskip}
1&1&0&0&1&1&77\,107.4757&2.056(-5)\\
1&1&2&0&1&2&77\,107.7671&1.542(-5)\\
1&1&2&0&1&1&77\,107.7671&5.140(-6)\\
1&1&1&0&1&1&77\,107.9023&5.140(-6)\\
1&1&1&0&1&0&77\,107.9023&6.854(-6)\\
1&1&1&0&1&2&77\,107.9023&8.568(-6)\\
1&2&2&0&1&1&77\,109.3248&1.542(-5)\\
1&2&2&0&1&2&77\,109.3248&5.141(-6)\\
1&2&3&0&1&2&77\,109.6162&2.056(-5)\\
1&2&1&0&1&0&77\,109.8104&1.142(-5)\\
1&2&1&0&1&2&77\,109.8104&5.712(-7)\\
1&2&1&0&1&1&77\,109.8104&8.568(-6)\\
1&0&1&0&1&2&77\,112.1085&1.142(-5)\\
1&0&1&0&1&0&77\,112.1085&2.285(-6)\\
1&0&1&0&1&1&77\,112.1085&6.855(-6)\\
 \noalign{\smallskip} \hline \noalign{\smallskip}
\end{tabular}
\end{table}
}
\onltab{9}{
\begin{table}[htbp]
 \caption{Hyperfine components and A$_{ul}$ Einstein spontaneous emission coefficients of the \jdub transition of N$_2$D$^+$. The frequency uncertainty is $\pm$ 2.1 kHz for all hyperfine components}\label{tabn2dp21}
 \centering
 \begin{tabular}{ccccccD{.}{.}{4}D{.}{.}{3}}

 \hline \noalign{\smallskip}
  $J'$ & $F_1'$ &  \multicolumn{2}{c}{$F' \rightarrow J$}  & $F_1$ & $F$ & \multicolumn{1}{c}{Frequency} & \multicolumn{1}{c}{A$_{ul}$} \cr
  &&&&&& \multicolumn{1}{c}{(MHz)}& \multicolumn{1}{c}{(s$^{-1}$)}\cr
 \noalign{\smallskip} \hline \noalign{\smallskip}
2&2&2&1&2&1&154\,214.8196& 7.402(-6)\\
2&2&2&1&2&3&154\,215.0138& 7.676(-6)\\
2&2&1&1&2&1&154\,215.1417& 3.701(-5)\\
2&2&3&1&2&3&154\,215.2619& 4.387(-5)\\
2&2&2&1&2&2&154\,215.3052& 3.427(-5)\\
2&1&1&1&0&1&154\,215.3991& 1.097(-4)\\
2&2&3&1&2&2&154\,215.5533& 5.483(-6)\\
2&1&2&1&0&1&154\,215.6021& 1.097(-4)\\
2&2&1&1&2&2&154\,215.6273& 1.234(-5)\\
2&1&0&1&0&1&154\,215.8617& 1.097(-4)\\
2&2&2&1&1&1&154\,216.7277& 1.110(-4)\\
2&3&3&1&2&3&154\,216.7920& 2.193(-5)\\
2&2&2&1&1&2&154\,216.8629& 3.701(-5)\\
2&2&1&1&1&1&154\,217.0498& 6.169(-5)\\
2&3&3&1&2&2&154\,217.0834& 1.755(-4)\\
2&3&2&1&2&1&154\,217.1055& 1.658(-4)\\
2&2&3&1&1&2&154\,217.1110& 1.481(-4)\\
2&3&4&1&2&3&154\,217.1807& 1.974(-4)\\
2&2&1&1&1&2&154\,217.1850& 4.113(-6)\\
2&3&2&1&2&3&154\,217.2998& 8.773(-7)\\
2&2&1&1&1&0&154\,217.4764& 8.225(-5)\\
2&3&2&1&2&2&154\,217.5912& 3.071(-5)\\
2&1&1&1&2&1&154\,217.6972& 1.371(-6)\\
2&1&2&1&2&1&154\,217.9002& 5.483(-8)\\
2&1&2&1&2&3&154\,218.0944& 4.606(-6)\\
2&1&0&1&2&1&154\,218.1598& 5.483(-6)\\
2&1&1&1&2&2&154\,218.1828& 4.113(-6)\\
2&1&2&1&2&2&154\,218.3858& 8.225(-7)\\
2&1&1&1&1&1&154\,219.6053& 2.056(-5)\\
2&1&1&1&1&2&154\,219.7405& 3.427(-5)\\
2&1&2&1&1&1&154\,219.8083& 2.056(-5)\\
2&1&2&1&1&2&154\,219.9435& 6.169(-5)\\
2&1&1&1&1&0&154\,220.0320& 2.742(-5)\\
2&1&0&1&1&1&154\,220.0679& 8.225(-5)\\
 \noalign{\smallskip} \hline \noalign{\smallskip}
\end{tabular}
\end{table}
}
\onltab{10}{
\begin{table}[htbp]
\vspace{1cm}
 \caption{Hyperfine components and A$_{ul}$ Einstein spontaneous emission coefficients of the \jtdb transition of N$_2$D$^+$. The frequency uncertainty is $\pm$ 6.2 kHz for all hyperfine components}\label{tabn2dp32}
 \centering
 \begin{tabular}{ccccccD{.}{.}{4}D{.}{.}{3}}

 \hline \noalign{\smallskip}
  $J'$ & $F_1'$ &  \multicolumn{2}{c}{$F' \rightarrow J$}  & $F_1$ & $F$ & \multicolumn{1}{c}{Frequency} & \multicolumn{1}{c}{A$_{ul}$} \cr
  &&&&&& \multicolumn{1}{c}{(MHz)}& \multicolumn{1}{c}{(s$^{-1}$)}\cr
 \noalign{\smallskip} \hline \noalign{\smallskip}
3&3&3&2&3&2&231\,319.4552& 6.294(-6)\\
3&3&3&2&3&4&231\,319.5743& 6.373(-6)\\
3&3&2&2&3&2&231\,319.8904& 7.049(-5)\\
3&3&4&2&3&4&231\,319.9411& 7.435(-5)\\
3&3&3&2&3&3&231\,319.9629& 6.664(-5)\\
3&3&4&2&3&3&231\,320.3297& 4.957(-6)\\
3&3&2&2&3&3&231\,320.3981& 8.812(-6)\\
3&2&2&2&1&2&231\,321.1993& 1.499(-4)\\
3&2&2&2&1&1&231\,321.4023& 4.497(-4)\\
3&2&1&2&1&0&231\,321.4445& 3.331(-4)\\
3&4&4&2&3&4&231\,321.4756& 4.461(-5)\\
3&3&3&2&2&3&231\,321.4930& 7.050(-5)\\
3&2&3&2&1&2&231\,321.5630& 5.996(-4)\\
3&2&1&2&1&2&231\,321.7041& 1.665(-5)\\
3&3&3&2&2&2&231\,321.7411& 5.640(-4)\\
3&3&2&2&2&1&231\,321.8543& 5.330(-4)\\
3&3&4&2&2&3&231\,321.8599& 6.345(-4)\\
3&4&4&2&3&3&231\,321.8643& 6.692(-4)\\
3&4&3&2&3&2&231\,321.8645& 6.555(-4)\\
3&2&1&2&1&1&231\,321.9071& 2.498(-4)\\
3&4&5&2&3&4&231\,321.9120& 7.138(-4)\\
3&3&2&2&2&3&231\,321.9283& 2.820(-6)\\
3&4&3&2&3&4&231\,321.9836& 9.104(-7)\\
3&2&2&2&3&2&231\,321.9940& 3.525(-7)\\
3&3&2&2&2&2&231\,322.1764& 9.870(-5)\\
3&2&3&2&3&2&231\,322.3577& 7.194(-9)\\
3&4&3&2&3&3&231\,322.3722& 5.736(-5)\\
3&2&3&2&3&4&231\,322.4768& 2.913(-6)\\
3&2&1&2&3&2&231\,322.4988& 3.172(-6)\\
3&2&2&2&3&3&231\,322.5017& 2.820(-6)\\
3&2&3&2&3&3&231\,322.8654& 2.518(-7)\\
3&2&2&2&2&1&231\,323.9578& 1.666(-5)\\
3&2&2&2&2&3&231\,324.0318& 1.727(-5)\\
3&2&2&2&2&2&231\,324.2799& 7.711(-5)\\
3&2&3&2&2&3&231\,324.3955& 9.870(-5)\\
3&2&1&2&2&1&231\,324.4626& 8.328(-5)\\
3&2&3&2&2&2&231\,324.6436& 1.234(-5)\\
3&2&1&2&2&2&231\,324.7847& 2.776(-5)\\
 \noalign{\smallskip} \hline \noalign{\smallskip}
\end{tabular}
\end{table}
}
\onltab{11}{
\begin{table}[htbp]
 \caption{Hyperfine components and A$_{ul}$ Einstein spontaneous emission coefficients of the \jqtb transition of N$_2$D$^+$. The frequency uncertainty is $\pm$ 25 kHz for all hyperfine components}\label{tabn2dp43}
 \centering
 \begin{tabular}{ccccccD{.}{.}{4}D{.}{.}{3}}

 \hline \noalign{\smallskip}
  $J'$ & $F_1'$ &  \multicolumn{2}{c}{$F' \rightarrow J$}  & $F_1$ & $F$ & \multicolumn{1}{c}{Frequency} & \multicolumn{1}{c}{A$_{ul}$} \cr
  &&&&&& \multicolumn{1}{c}{(MHz)}& \multicolumn{1}{c}{(s$^{-1}$)}\cr
 \noalign{\smallskip} \hline \noalign{\smallskip}
4&4&4&3&4&3&308\,419.723& 5.330(-6)\\
4&4&4&3&4&5&308\,419.794& 5.361(-6)\\
4&4&3&3&4&3&308\,420.188& 1.028(-4)\\
4&4&5&3&4&5&308\,420.218& 1.053(-4)\\
4&4&4&3&4&4&308\,420.231& 9.896(-5)\\
4&4&5&3&4&4&308\,420.655& 4.386(-6)\\
4&4&3&3&4&4&308\,420.696& 6.853(-6)\\
4&3&3&3&2&3&308\,421.627& 1.790(-4)\\
4&5&5&3&4&5&308\,421.764& 7.018(-5)\\
4&4&4&3&3&4&308\,421.765& 1.028(-4)\\
4&3&3&3&2&2&308\,421.991& 1.432(-3)\\
4&3&2&3&2&1&308\,421.993& 1.353(-3)\\
4&3&4&3&2&3&308\,422.056& 1.611(-3)\\
4&3&3&3&4&3&308\,422.120& 1.399(-7)\\
4&4&4&3&3&3&308\,422.132& 1.542(-3)\\
4&3&2&3&2&3&308\,422.134& 7.161(-6)\\
4&4&3&3&3&2&308\,422.162& 1.511(-3)\\
4&4&5&3&3&4&308\,422.189& 1.645(-3)\\
4&5&4&3&4&3&308\,422.195& 1.668(-3)\\
4&5&5&3&4&4&308\,422.200& 1.684(-3)\\
4&5&6&3&4&5&308\,422.230& 1.754(-3)\\
4&4&3&3&3&4&308\,422.231& 2.098(-6)\\
4&5&4&3&4&5&308\,422.266& 8.664(-7)\\
4&3&2&3&2&2&308\,422.498& 2.506(-4)\\
4&3&4&3&4&3&308\,422.549& 1.727(-9)\\
4&4&3&3&3&3&308\,422.598& 1.322(-4)\\
4&3&4&3&4&5&308\,422.621& 2.127(-6)\\
4&3&2&3&4&3&308\,422.628& 2.238(-6)\\
4&3&3&3&4&4&308\,422.628& 2.098(-6)\\
4&5&4&3&4&4&308\,422.703& 8.577(-5)\\
4&3&4&3&4&4&308\,423.057& 1.088(-7)\\
4&3&3&3&3&2&308\,424.094& 1.119(-5)\\
4&3&3&3&3&4&308\,424.163& 1.133(-5)\\
4&3&3&3&3&3&308\,424.530& 1.185(-4)\\
4&3&4&3&3&4&308\,424.592& 1.322(-4)\\
4&3&2&3&3&2&308\,424.602& 1.253(-4)\\
4&3&4&3&3&3&308\,424.959& 8.812(-6)\\
4&3&2&3&3&3&308\,425.037& 1.567(-5)\\
 \noalign{\smallskip} \hline \noalign{\smallskip}
\end{tabular}
\end{table}
}
\onltab{12}{
\begin{table}[htbp]
 \caption{Hyperfine components and A$_{ul}$ Einstein spontaneous emission coefficients of the \jcqb transition of N$_2$D$^+$. The frequency uncertainty is $\pm$ 58 kHz for all hyperfine components}\label{tabn2dp54}
 \centering
 \begin{tabular}{ccccccD{.}{.}{4}D{.}{.}{3}}

 \hline \noalign{\smallskip}
  $J'$ & $F_1'$ &  \multicolumn{2}{c}{$F' \rightarrow J$}  & $F_1$ & $F$ & \multicolumn{1}{c}{Frequency} & \multicolumn{1}{c}{A$_{ul}$} \cr
  &&&&&& \multicolumn{1}{c}{(MHz)}& \multicolumn{1}{c}{(s$^{-1}$)}\cr
 \noalign{\smallskip} \hline \noalign{\smallskip}
5&5&5&4&5&4&385\,514.088& 4.587(-6)\\
5&5&5&4&5&6&385\,514.125& 4.601(-6)\\
5&5&4&4&5&4&385\,514.562& 1.346(-4)\\
5&5&6&4&5&6&385\,514.584& 1.363(-4)\\
5&5&5&4&5&5&385\,514.591& 1.310(-4)\\
5&5&6&4&5&5&385\,515.050& 3.894(-6)\\
5&5&4&4&5&5&385\,515.065& 5.607(-6)\\
5&4&4&4&3&4&385\,516.051& 2.082(-4)\\
5&5&5&4&4&5&385\,516.136& 1.346(-4)\\
5&6&6&4&5&6&385\,516.141& 9.734(-5)\\
5&4&4&4&5&4&385\,516.406& 6.922(-8)\\
5&4&3&4&3&2&385\,516.474& 3.059(-3)\\
5&4&4&4&3&3&385\,516.480& 3.123(-3)\\
5&4&5&4&3&4&385\,516.516& 3.331(-3)\\
5&4&3&4&3&4&385\,516.553& 4.249(-6)\\
5&5&5&4&4&4&385\,516.560& 3.230(-3)\\
5&5&4&4&4&3&385\,516.568& 3.198(-3)\\
5&5&6&4&4&5&385\,516.595& 3.364(-3)\\
5&6&5&4&5&4&385\,516.599& 3.388(-3)\\
5&6&6&4&5&5&385\,516.607& 3.407(-3)\\
5&5&4&4&4&5&385\,516.610& 1.661(-6)\\
5&6&7&4&5&6&385\,516.627& 3.504(-3)\\
5&6&5&4&5&6&385\,516.636& 8.045(-7)\\
5&4&5&4&5&4&385\,516.871& 5.721e-10\\
5&4&3&4&5&4&385\,516.907& 1.731(-6)\\
5&4&5&4&5&6&385\,516.908& 1.673(-6)\\
5&4&4&4&5&5&385\,516.908& 1.661(-6)\\
5&4&3&4&3&3&385\,516.982& 2.677(-4)\\
5&5&4&4&4&4&385\,517.034& 1.645(-4)\\
5&6&5&4&5&5&385\,517.102& 1.150(-4)\\
5&4&5&4&5&5&385\,517.373& 5.663(-8)\\
5&4&4&4&4&3&385\,518.412& 8.328(-6)\\
5&4&4&4&4&5&385\,518.454& 8.376(-6)\\
5&4&4&4&4&4&385\,518.878& 1.546(-4)\\
5&4&3&4&4&3&385\,518.914& 1.606(-4)\\
5&4&5&4&4&5&385\,518.919& 1.645(-4)\\
5&4&5&4&4&4&385\,519.343& 6.853(-6)\\
5&4&3&4&4&4&385\,519.379& 1.071(-5)\\
 \noalign{\smallskip} \hline \noalign{\smallskip}
\end{tabular}
\end{table}
}
\onltab{13}{
\begin{table}[htbp]
 \caption{Hyperfine components and A$_{ul}$ Einstein spontaneous emission coefficients of the \jscb transition of N$_2$D$^+$. The frequency uncertainty is $\pm$ 0.11 MHz for all hyperfine components}\label{tabn2dp65}
 \centering
 \begin{tabular}{ccccccD{.}{.}{4}D{.}{.}{3}}

 \hline \noalign{\smallskip}
  $J'$ & $F_1'$ &  \multicolumn{2}{c}{$F' \rightarrow J$}  & $F_1$ & $F$ & \multicolumn{1}{c}{Frequency} & \multicolumn{1}{c}{A$_{ul}$} \cr
  &&&&&& \multicolumn{1}{c}{(MHz)}& \multicolumn{1}{c}{(s$^{-1}$)}\cr
 \noalign{\smallskip} \hline \noalign{\smallskip}
6&6&6&5&6&5&462\,601.05& 4.014(-6)\\
6&6&6&5&6&7&462\,601.06& 4.021(-6)\\
6&6&5&5&6&5&462\,601.53& 1.660(-4)\\
6&6&7&5&6&7&462\,601.54& 1.673(-4)\\
6&6&6&5&6&6&462\,601.55& 1.627(-4)\\
6&6&5&5&6&6&462\,602.02& 4.744(-6)\\
6&6&7&5&6&6&462\,602.03& 3.485(-6)\\
6&5&5&5&4&5&462\,603.04& 2.378(-4)\\
6&6&6&5&5&6&462\,603.10& 1.660(-4)\\
6&7&7&5&6&7&462\,603.11& 1.255(-4)\\
6&5&5&5&6&5&462\,603.32& 3.920(-8)\\
6&5&4&5&4&3&462\,603.50& 5.651(-3)\\
6&5&5&5&4&4&462\,603.51& 5.707(-3)\\
6&5&6&5&4&5&462\,603.53& 5.945(-3)\\
6&5&4&5&4&5&462\,603.54& 2.936(-6)\\
6&6&5&5&5&4&462\,603.56& 5.780(-3)\\
6&6&6&5&5&5&462\,603.56& 5.811(-3)\\
6&6&5&5&5&6&462\,603.58& 1.372(-6)\\
6&6&7&5&5&6&462\,603.59& 5.977(-3)\\
6&7&6&5&6&5&462\,603.59& 6.002(-3)\\
6&7&7&5&6&6&462\,603.60& 6.022(-3)\\
6&7&6&5&6&7&462\,603.60& 7.424(-7)\\
6&7&8&5&6&7&462\,603.61& 6.148(-3)\\
6&5&6&5&6&5&462\,603.81& 2.320e-10\\
6&5&4&5&6&5&462\,603.81& 1.411(-6)\\
6&5&5&5&6&6&462\,603.81& 1.372(-6)\\
6&5&6&5&6&7&462\,603.81& 1.378(-6)\\
6&5&4&5&4&4&462\,604.00& 2.906(-4)\\
6&6&5&5&5&5&462\,604.04& 1.962(-4)\\
6&7&6&5&6&6&462\,604.09& 1.448(-4)\\
6&5&6&5&6&6&462\,604.30& 3.317(-8)\\
6&5&5&5&5&4&462\,605.35& 6.605(-6)\\
6&5&5&5&5&6&462\,605.37& 6.626(-6)\\
6&5&5&5&5&5&462\,605.83& 1.886(-4)\\
6&5&4&5&5&4&462\,605.85& 1.938(-4)\\
6&5&6&5&5&6&462\,605.86& 1.962(-4)\\
6&5&6&5&5&5&462\,606.32& 5.606(-6)\\
6&5&4&5&5&5&462\,606.32& 8.073(-6)\\
\noalign{\smallskip} \hline \noalign{\smallskip}
\end{tabular}
\end{table}
}
\end{document}